\documentclass{ws-procs9x6-cpt19}
\begin{document}

\newcommand{\refeq}[1]{(\ref{#1})}
\def\etal {{\it et al.}}

\title{Prospects for Lorentz-Violation Searches\\ 
at the LHC and Future Colliders}

\author{N.\ Chanon, A.\ Carle, and S. Perri\`es}

\address{Universit\'e de Lyon,
Universit\'e Claude Bernard Lyon 1,\\
CNRS-IN2P3,
Institut de Physique Nucl\'eaire de Lyon,\\
Villeurbanne 69622, France}

\begin{abstract}
Hadron colliders are providing a unique opportunity 
for testing Lorentz invariance and CPT symmetry 
at high energy and in a laboratory. 
A first measurement in the top-quark sector 
was performed at the Tevatron. 
We present here prospective studies for testing Lorentz invariance 
in top-quark pair production at the LHC and future colliders. 
The $b$-quark sector was investigated recently at LHCb. 
Eventually, 
new bounds on photon parameters 
can be extracted from the observation of TeV photons 
at the LHC. 
We will conclude by highlighting other opportunities 
provided by hadron colliders.
\end{abstract}

\bodymatter

\section{Testing Lorentz invariance at hadron colliders}

Lorentz invariance is a fundamental symmetry of the Standard Model, 
not necessarily expected to be conserved in theories of quantum gravity, 
such as string theories.\cite{Kostelecky:1988zi}
Hadron colliders offer the opportunity 
to probe Lorentz invariance and CPT symmetry 
(CPT breaking implies violation of Lorentz invariance under mild assumptions) 
with elementary particles at high energy.
The LHC provides proton--proton collisions 
at a center-of-mass energy of $13\,$TeV (Run 2). 
Although the scale of Lorentz- or CPT-symmetry breaking 
is sometimes expected to lie at the Planck mass, 
the extra-dimensions paradigm could lower the scale of quantum gravity. 
Possible remnants of Lorentz-symmetry breaking can be analyzed 
using signatures proposed within the framework of an effective field theory, 
the Standard-Model Extension (SME).\cite{Colladay:1996iz}
Among the processes occurring with the highest cross sections in $p$--$p$ collisions, 
and where such signatures could be searched for, 
one can list jet production, 
photon production, 
$W$ or $Z$ production, 
and $t\bar{t}$ production. 
We will focus in these proceedings on photon and top-quark production 
as a probes for Lorentz-violation (LV) searches.

\section{Probing LV in the top-quark sector at hadron colliders}

\subsection{Searches for LV with $t\bar{t}$ production}

The top-quark sector in the SME is weakly constrained 
with only one direct measurement, 
performed using the D0 detector at the Tevatron\cite{Abazov:2012iu} 
via observation of $t\bar{t}$ events. 
The obtained results are compatible with no LV 
with an absolute uncertainty of 10\%. 

The top-quark lagrangian in the SME 
contains both CPT-even and CPT-odd LV contributions. 
Since top and antitop CPT-violating corrections cancel in $t\bar{t}$ production, 
we will focus on the CPT-even LV contribution. 
The SME introduces the LV $c_{\mu\nu}$ coefficients, 
modifying the top-quark propagator, 
the $Wtb$ vertex and the top--gluon coupling. 
The squared matrix elements for top-quark production and decay 
including these corrections are known\cite{Berger:2015yha} 
at leading order in perturbative QCD.

The LHC is a top-quark factory. 
From the Tevatron to the LHC, 
the center-of-mass energy has increased 
from $1.96\,$TeV to $13\,$TeV. 
The cross section for $t\bar{t}$ production 
has increased by a factor $\approx$115 
owing to the gluon luminosity in the proton. 
Furthermore, 
the recorded luminosity at LHC Run 2 
is about $150\,$fb$^{-1}$, 
while D0 results\cite{Abazov:2012iu} 
were obtained analyzing $5.3\,$fb$^{-1}$. 
An even higher number of $t\bar{t}$ events 
will be produced at future hadron colliders, 
such as the HL-LHC ($3\,$ab$^{-1}$ at $14\,$TeV) 
and the HE-LHC or FCC-hh options 
($15\,$ab$^{-1}$ at $27\,$TeV and $100\,$TeV, respectively). 

We compute the change in $t\bar{t}$ cross section in the SME framework. 
The SME coefficients are constant in a given inertial reference frame 
taken by convention to be the Sun-centered frame. 
The rotation of the Earth around its axis 
induces a sinusoidal modulation of the $t\bar{t}$ cross section 
as a function of time. 
We use the location of ATLAS or CMS experiments 
(both detectors are located at opposite azimuth in the LHC ring 
and ``see'' the same cross section). 
The experiments are sensitive to the same $c_{\mu\nu}$ 
coefficients as those measured at D0. 
It is found that the amplitude of the induced modulation 
is growing with the center-of-mass energy.

We evaluate the expected precision on $c_{\mu\nu}$ for several benchmarks: 
D0, LHC Run 2, HL-LHC, HE-LHC, FCC.\cite{CarleChanonPerries}
From an absolute precision of 0.1 on the coefficients measured at D0, 
we find that the expected sensitivity would improve
by a factor $10^{2}$--$10^{3}$ at LHC Run 2, 
the exact value depending on the $c_{\mu\nu}$ benchmark. 
An additional improvement by another factor $10^2$ 
is expected at future colliders like the FCC. 
As a summary, 
there is great potential for precision measurement of top-quark $c_{\mu\nu}$ coefficients
at present and future hadron colliders.

\subsection{Searches for CPT violation with single-top production}

The single-top-quark production is sensitive to CPT violation. 
By comparing the rates for single-top and single-antitop production 
as a function of time, 
the $b_{\mu}$ coefficient in the SME could be measured.\cite{Berger:2015yha} 
However, 
the measurement is challenging 
owing to a huge $t\bar{t}$ background. 
While the $s$-channel single-top production 
remains to be observed at the LHC, 
the $t$-channel or $tW$-channel could be investigated.

\section{A test of CPT violation through $B_{s}$ oscillations at LHCb}

The first test of CPT violation at the LHC 
in the context of the SME was recently performed at LHCb
using $B^{0}$ and $B_{s}$ neutral-meson oscillations.\cite{Aaij:2016mos}
The tiny mass difference between the $B_{s}$ 
and its antiparticle is used to achieve excellent precision 
on the $\Delta a_{\mu}$ SME coefficients for $b$ quarks. 
The analysis, 
performed as a function of sidereal time, 
achieved a precision of 10$^{-14}\,$GeV, 
a factor $10^{2}$ improvement relative to previous measurements. 

While this measurement was performed 
with the LHC data collected at 7 and $8\,$TeV, 
it could already be improved 
by using the $13\,$TeV data of LHC Run 2. 
There is also potential for improvement by a factor of three 
of the limits on $c$-quark SME coefficients 
by analyzing the $D^0 \rightarrow K^{-} \pi^{+}$ decay.\cite{vanTilburg:2014dka}

\section{Direct photons and constraints on the SME}

LV in the photon sector can, 
for example, 
be searched for with vacuum birefringence, 
polarization-independent anisotropies of the phase-speed of light, 
or an isotropic shift in the phase-speed of light.\cite{Hohensee:2008xz} 
The latter is controlled by the $\tilde{\kappa}_{tr}$ coefficient 
in the SME. 
A positive $\tilde{\kappa}_{tr}$ 
leads to Cherenkov radiation of charged particles in vacuum, 
and a negative $\tilde{\kappa}_{tr}$ leads to photon decay 
into a fermion--antifermion pair in vacuum;
both of these processes are forbidden in the Standard Model. 
The smaller $-\tilde{\kappa}_{tr}$, 
the higher the threshold energy 
at which photon decay occurs, 
and the longer photons travel before decaying. 
The interplay between $\tilde{\kappa}_{tr}$ 
and the fermion $c_{\mu\nu}$ coefficients implies 
that only the quantity $\tilde{\kappa}_{tr} -(4/3)c_{TT}$ 
can actually be measured. 

Today's best limits on this coefficient 
are obtained using photons of astrophysical origin. 
Measurements of high-energy photon showers on Earth imply 
that the photon traveled a large distance without decaying, 
hence the threshold for photon decay was not reached,
and the photon must have been below the LV pair-production threshold. 
Bounds on $\tilde{\kappa}_{tr} -(4/3)c_{TT}$ 
can then be extracted 
(assuming, for instance, electrons as decay products).

The D0 experiment at the Tevatron 
observed photons up to $340.5\,$GeV. 
By assuming conservatively 
that the threshold for photon decay is $300\,$GeV, 
and assuming the process $\gamma \rightarrow e^{+}e^{-}$, 
a bound  from direct photon production at hadron colliders was set:\cite{Hohensee:2008xz} $\tilde{\kappa}_{tr} -(4/3)c_{TT} > -5.8 \times 10^{-12}$. 

At the LHC, 
the ATLAS experiment recorded photons 
with a transverse energy above $1.1\,$TeV with a minimum pseudorapidity $|\eta|>0.6$,\cite{Aaboud:2017cbm} 
such that we can derive a threshold energy of $1.3\,$TeV. 
A new bound can then be extracted: 
$\tilde{\kappa}_{tr} -(4/3)c_{TT} > -3.1 \times 10^{-13}$, 
an improvement by almost a factor 20 
relative to the previous collider bound. 
However, 
since the energy threshold is higher, 
the photon would travel on average a longer distance 
than at D0 before decaying, 
and in some cases the decay could still be reconstructed as a converted photon. 
This idea would deserve a more detailed analysis.

\section{Conclusions}

In these proceedings, 
we presented an overview of possible signatures 
for violations of Lorentz invariance and CPT symmetry at hadron colliders 
involving top quarks, 
neutral-meson oscillation, 
and direct photon production. 
Hadron colliders present also other opportunities: 
with the high cross section for QCD jet production, 
or $W$ and $Z$ production, 
interesting new searches could be performed. 

\section*{Acknowledgments}
The author is thankful to the organizers for the invitation and for the fruitful discussions at the meeting.

\end{document}